\documentstyle[psfig,floats,aps,aipbook]{revtex}
\psfigurepath {.}
\pssilent
\newcommand{\figside}[3]{
\begin{figure}[t]
\begin{center}
\begin{tabular}{lp{6cm}}
 \raisebox{-4cm}
{#1} & {#2}
\end{tabular}
\end{center}
\refstepcounter{figure} #3
\vspace{-8mm}
\end{figure}
}

\newcommand{\ignore}[1]{ }

\def\kms{\hbox{km~s$^{-1}$}}
\def\mathnew{\mathsurround=0pt}
\def\simov#1#2{\lower .5pt\vbox{\baselineskip0pt \lineskip-.5pt
        \ialign{$\mathnew#1\hfil##\hfil$\crcr#2\crcr\sim\crcr}}}
\def\simgreat{\mathrel{\mathpalette\simov >}}
\def\simless{\mathrel{\mathpalette\simov <}}

\def\cm{\hbox{cm}}
\def\s{\hbox{s}}
\def\erg{\hbox{erg}}
\def\G{\hbox{G}}

\def\MeV{\hbox{MeV}}
\def\g{\hbox{g}}
\def\yr{\hbox{yr}}
\def\dyne{\hbox{dyne}}

\def\sun{\odot}

\def\kpc{\hbox{kpc}}

\begin{document}

\title{Galactic Models of Gamma-Ray Bursts}

\author{Donald Q. Lamb$^*$, Tomasz Bulik,$^*$ and Paolo S. Coppi$^\dagger$}

\address{$^*$Department of Astronomy and Astrophysics, University of Chicago\\
5640 South Ellis Avenue, Chicago, IL 60637\\
$^\dagger$Department of Astronomy, Yale University\\
P.O. Box 208101, New Haven, CT  06520}

\maketitle
\begin{abstract}
We describe observational evidence and theoretical calculations which
support the high velocity neutron star model of gamma-ray bursts.  We
estimate the  energetic requirements in this model, and discuss possible energy
sources. we also consider radiative processes involved in the bursts.
\end{abstract}

\vspace*{-3.6mm}
\section*{Introduction}

\vspace*{-3.1mm}
Gamma-ray bursts (GRBs) continue to confound astrophysicists nearly a
quarter century after their discovery \cite{KSO73}.  Before the
launch of CGRO, most scientists thought that GRBs came from magnetic
neutron stars residing in a thick disk (having a scale height of up to
$\sim$ 2 kpc) in the Milky Way \cite{HigLin90,Hard91}.  The data
gathered by BATSE showed the existence of a rollover in the cumulative
brightness distribution of GRBs and that the sky distribution of even
faint GRBs is consistent with isotropy \cite{Meegan92,Briggs95}.
This rules out a thick Galactic disk source population, with
\cite{Smith94} or without \cite{MaoP92} spiral arms.

Consequently, the primary impact of the BATSE results has been to
intensify debate about whether the bursts are Galactic or cosmological
in origin.  Galactic models attribute the bursts primarily to
high-velocity neutron stars in a Galactic corona, which must
extend one sixth or more of the distance to Andromeda ($d_{\rm M31} \sim
690$ kpc) in order to avoid any discernible anisotropy
\cite{Hak94,Hartmann94}.  Cosmological models place the GRB sources at
distances $d \sim 1 - 3$ Gpc, corresponding to redshifts $z\sim 0.3 -
1$.  A source population at such large distances naturally produces an
isotropic distribution of bursts on the sky, and the expansion of the
universe or source evolution can reproduce the observed rollover in the
cumulative brightness distribution \cite{Fen93}.

Within the context of this workshop, we focus on Galactic corona models
involving high velocity neutron stars.
A recent discussion of cosmological models may be found in, e.g., Blaes
\cite{Blaes94}.

\section*{High Velocity Neutron Stars}

\vspace*{-3.1mm}
Only a few years ago scientists thought that neutron stars had
velocities of 100 - 200 km s$^{-1}$ \cite{LAS82}.  But recent studies
show \cite{LyneLori94,FrailGW94} that as much as 50\% of neutron stars
have velocities $v > 800$ km s$^{-1}$.  These velocities are so high
that these neutron stars escape from the Galaxy and produce a distant,
previously unknown Galactic "corona."

The evidence that many neutron stars have high velocities comes from
two independent directions.  In the first case, long-wavelength radio
observations have discovered that many young radio pulsars are
associated with young ($t_{\rm age} < 10^4$ yrs) supernova remnants
\cite{FrailGW94}.  Sometimes the young pulsar lies within
the shell-like supernova remnant; sometimes it is passing through the
shell, as the spectacular radio image of the "duck" supernova remnant
and pulsar PSR1757-24 reveals; and sometimes the young pulsar is
associated only with a comet-like "plerion," or filled remnant.  In
every case the pulsar lies far from the center of the remnant.  These
offsets imply median transverse velocities $\sim 500$ km s$^{-1}$, with
$\sim$ 1/3 of the neutron stars having transverse velocities $> 1000$
km s$^{-1}$  \cite{FrailGW94}.

In the second case, a new model for the electron density in the Milky
Way and a greater understanding of an important observational bias that
affects the determination of pulsar velocities has dramatically
increased the velocities inferred for older pulsars.  The new electron
density model shows that the distance to, and therefore the transverse
velocity of, nearby pulsars was underestimated by about a factor of two
in previous models \cite{TayCordes93}.  The observational bias that
affects the determination of pulsar velocities arises because young
radio pulsars are born close to the Galactic plane, and move rapidly
away from it if their velocity is high.  After some time, the pulsars
that remain within detectable range are mostly those with small
velocities.  The strength of the bias is illustrated by the fact that
the mean of the distribution of transverse velocities is $345 \pm 70$
km s$^{-1}$ for pulsars with spindown ages $\tau <$ 3 Myr, whereas it
is $105 \pm 25$ km s$^{-1}$ for pulsars with $\tau \simgreat$ 70 Myr
\cite{LyneLori94}.

Recent studies that incorporate these discoveries yield median neutron
star total velocities $\langle v \rangle_{\rm median} \sim 600$ km
s$^{-1}$, with as many as half of all neutron stars having  velocities
$v > 800$ km s$^{-1}$ \cite{LyneLori94,Chernoff95}.  These results have
revolutionized our understanding of the spatial distribution of neutron
stars in the Galaxy.  Since the escape velocity from the Milky Way is
$\approx 500$ km s$^{-1}$ in the solar neighborhood and $\approx 600$
km s$^{-1}$ in the Galactic bulge, where most neutron stars are born,
all of these high velocity neutron stars will escape from the Milky
Way.  They form a distant, previously unknown "corona" around the Milky
Way.  This distant corona contains an ample population of sources which
appear isotropic when viewed from the Earth.

\section*{The Galactic Corona}

\vspace*{-3.1mm}
Prior to the launch of CGRO, many scientists believed it likely that
gamma-ray bursts came from a thick Galactic disk.  But, while a
Galactic disk population was the most conservative and perhaps the most
popular model \cite{HigLin90,Hard91}, extended halo populations have
also had a long and illustrious history (see, e.g.,
\cite{Fish79,Jenn82,Jenn80,Shklov85}).
 What did exist was a consensus that
gamma-ray bursts come from magnetic neutron stars in the Galaxy.  There
were many reasons for this, several of which we describe below.

Following the discovery by BATSE that the faint bursts are distributed
isotropically on the sky, Galactic halo and corona models found new
flavor (see, e.g., \cite{Hartmann94,Brainerd92,DT92,LiDer:92,SmithLamb93})
 as an
attractive way of reconciling all of the evidence about GRBs  which favors
Galactic neutron stars with isotropy.  However,
these models were considered somewhat ad hoc, particularly by advocates
of cosmological models, because no means of producing large numbers of
neutron stars in an extended Galactic halo was known [see, e.g.
\cite{Pac93}].  .

Consequently, the debate about whether GRBs are Galactic or
cosmological in origin was characterized as one between those who
advocated objects which we know produce burst-like phenomena (high
velocity neutron stars; see below) but which were not known to have the
necessary spatial distribution (extended Galactic halo) {\it vs.} those
who advocated objects which we do not know can produce burst-like
phenomena (e.g., coalescing neutron star binaries or failed supernovae)
but were known to have the necessary spatial distribution
(cosmological).

The subsequent discovery that many neutron stars have velocities high
enough to escape from the Milky Way
has given   models a tremendous boost.  Nevertheless, these models
 must answer several important questions:
\begin{itemize}

\item[\rm $\bullet$] Can a   Galactic corona of high velocity neutron
stars account for the isotropic sky distribution and the rollover in
the brightness distribution of GRBs seen by BATSE?

\item[\rm $\bullet$] Why do only high velocity neutron stars produce
GRBs?

\item[\rm $\bullet$] Are GRBs beamed along the direction of
motion of the neutron star or, if not, why is bursting activity
delayed?

\item[\rm $\bullet$] Are there energy  sources   sufficient to power
GRBs in such a model?

\item[\rm $\bullet$] Can the energy needed be
 released over 500 Myr or more?

\item[\rm $\bullet$] Can cyclotron lines formed in regions where the magnetic
field is $\sim 2\times 10^{12} - 10^{13}$ G be produced by magnetic
neutron stars in GRBs with luminosities $L_{\rm burst } \sim 10^{41}
- 10^{43}$ erg s$^{-1}$?
\end{itemize}
We consider each of these questions below.

\subsection*{Ingredients in High Velocity Neutron Star Models}

\begin{figure}[t]
\begin{tabular}{lr}
 {\psfig{file=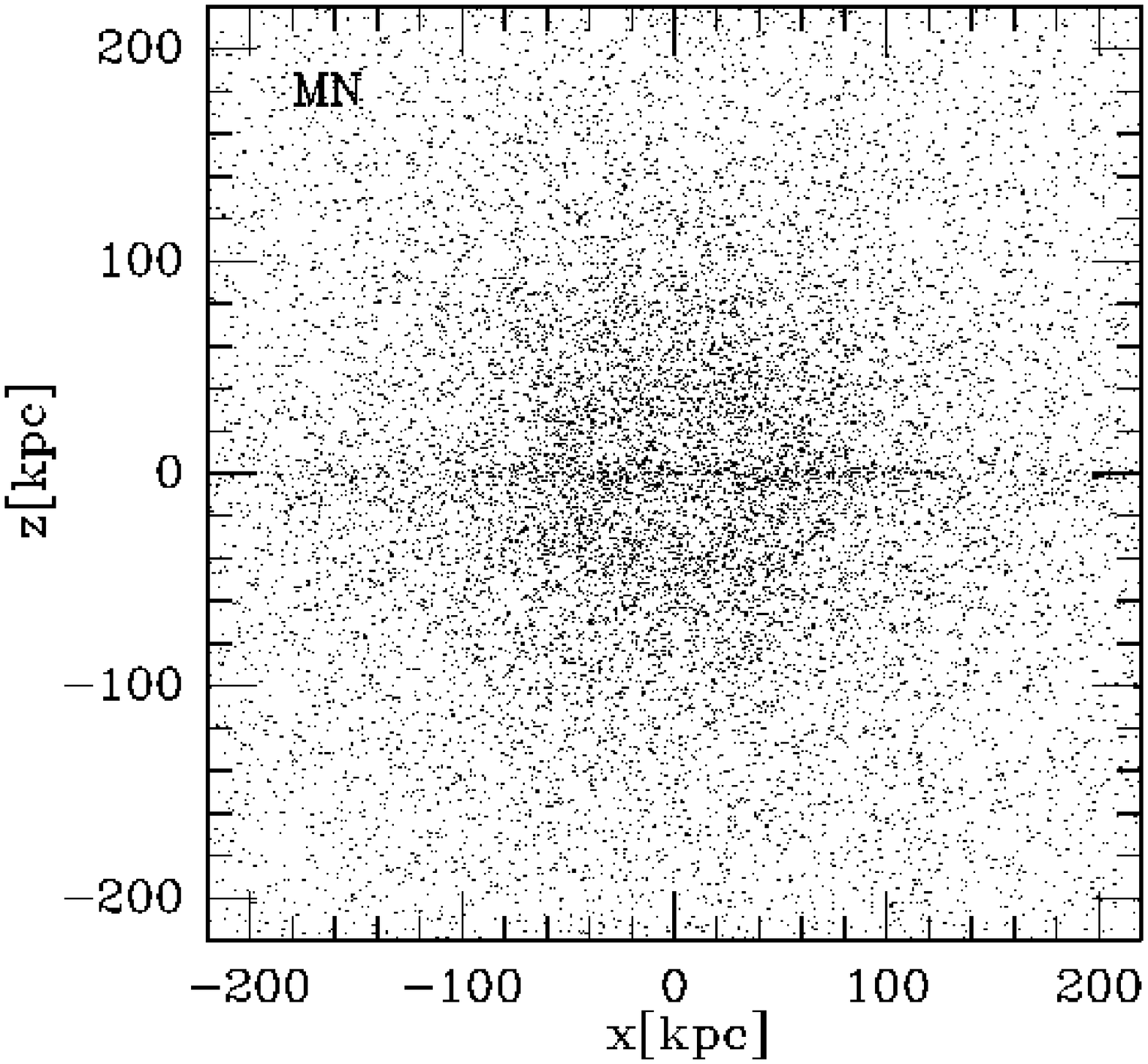,width=5.5cm,angle=-90}} &
 {\psfig{file=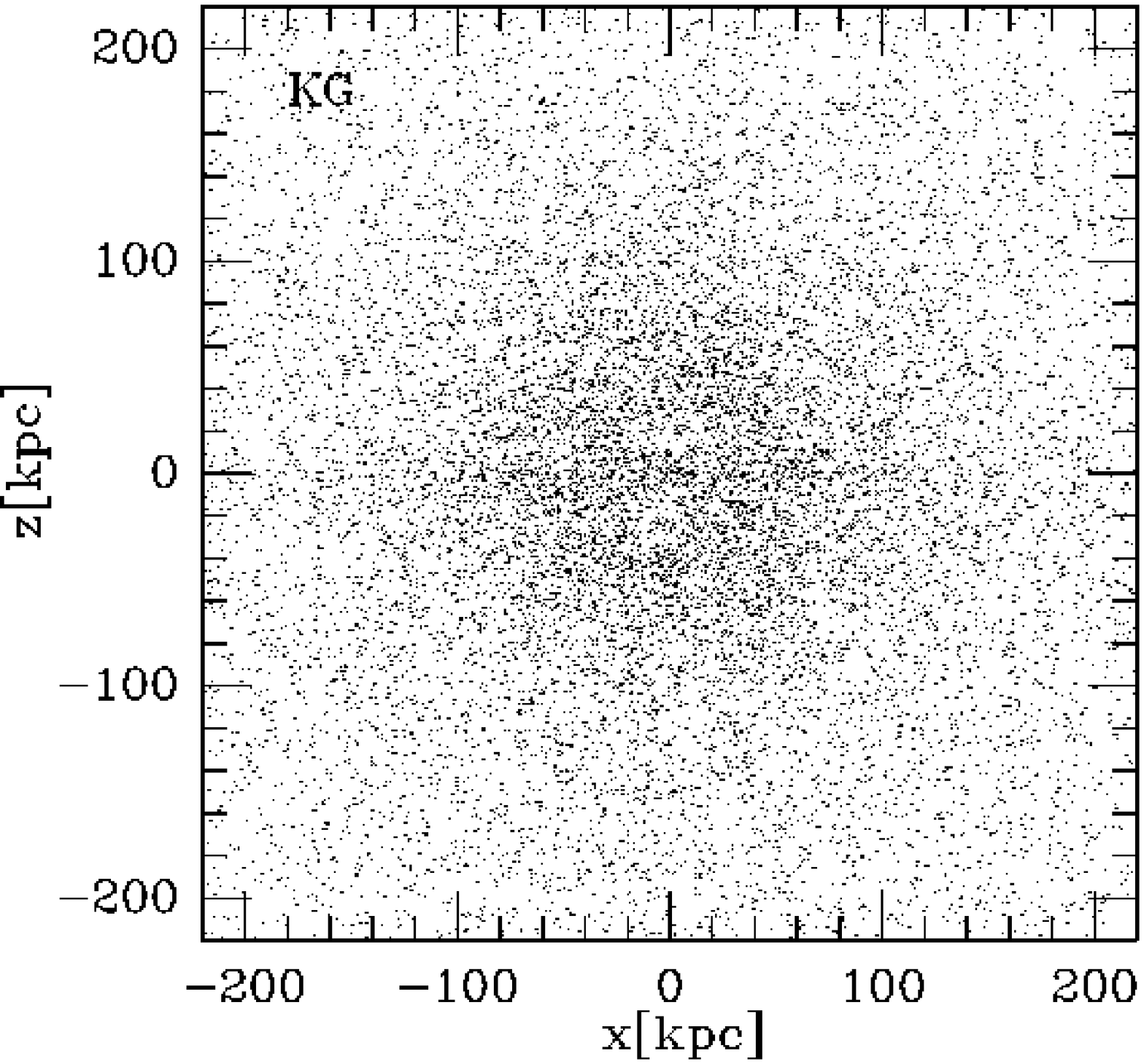,width=5.5cm,angle=-90}}
\end{tabular}
\caption{Distribution of neutron stars with an initial kick velocity of
$1000$~\kms\ found using the Miyamoto and Nagai (1975) potential (left panel)
  and using the Kuijken and Gilmore (1989) potential (right panel). Note
the increased concentration of stars in an extended disk due to the
focusing described in the text.}\vspace{-5mm}
\end{figure}

 \vspace*{-3.1mm}
We have calculated detailed models of the spatial distribution expected
for a population of high-velocity neutron stars born in the Galactic
disk and moving in a Galactic potential that includes the bulge, disk,
and a dark matter halo.  All earlier studies of which we are aware that
included these components of the gravitational potential employed the potential
given by Miyamoto and Nagai  \cite{MN75}. 
For studies of high-velocity neutron stars, it is
essential to use a potential that is realistic out to very large
distances.  We find that  the Miyamoto and Nagai \cite{MN75}
implies the existence
of an extended disk, far beyond the observed Galactic disk.
Approximately $\approx$ 20\% of the mass lies outside $r = 20$ kpc,
whereas in more realistic exponential disks less than 4\% of the mass
lies outside $20$~kpc.

  The
focusing effect of such a disk can be seen by calculating the ratio of
the $z$ components of the force using the Miyamoto and Nagai \cite{MN75}
potential to the force due to a point mass for $r\gg a, b, z$:
\begin{equation}
{F_z^{MN}\over F_z^{PM}} = 1 + {a\over(b^2 +z^2)^{1/2}},
\end{equation}
where $a,b$ are parameters describing the Miyamoto and Nagai potential.
Typically $a\approx 4$~kpc, $b\approx 0.2$~kpc, so that
$F_z^{MN}/F_z^{PM} \rightarrow \; \approx 20$ as $z \rightarrow 0$. The
 use of the Miyamoto and Nagai  \cite{MN75}
potential
distorts the orbits of neutron stars whose initial velocity vectors lie
in or near the plane of the disk, and leads to an anisotropic spatial
distribution that is entirely an artifact of the unrealistic disk
potential (see Figure 1).

  We therefore use the mass distribution and potential given by
Kuijken and Gilmore \cite{KG:89}. Details of our calculation are given
 in \cite{BulLam95a}.

\begin{figure}[t]
\begin{center}
\begin{tabular}{lr}
{\psfig{file=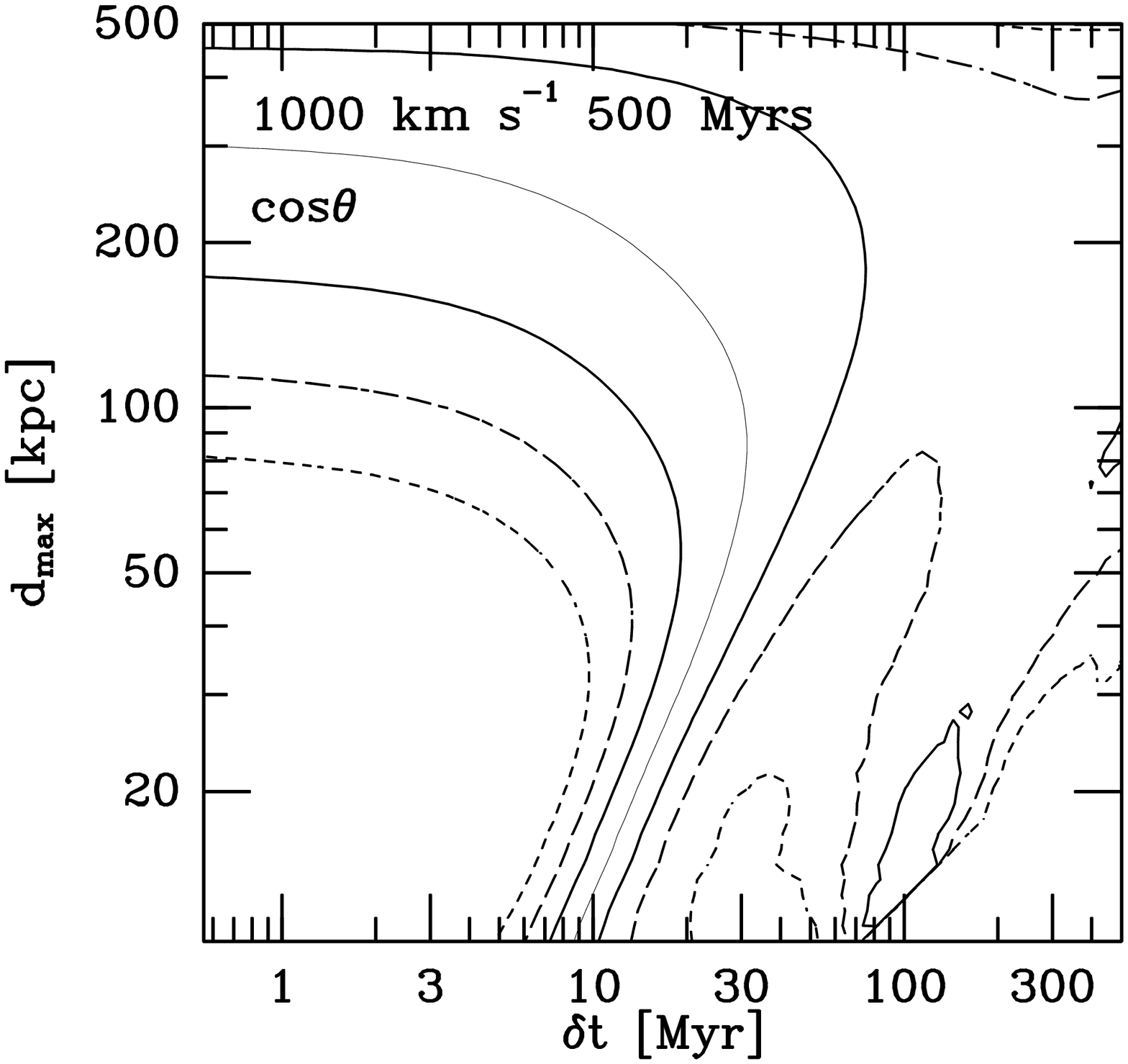,width=5.5cm,angle=-90}} &
{\psfig{file=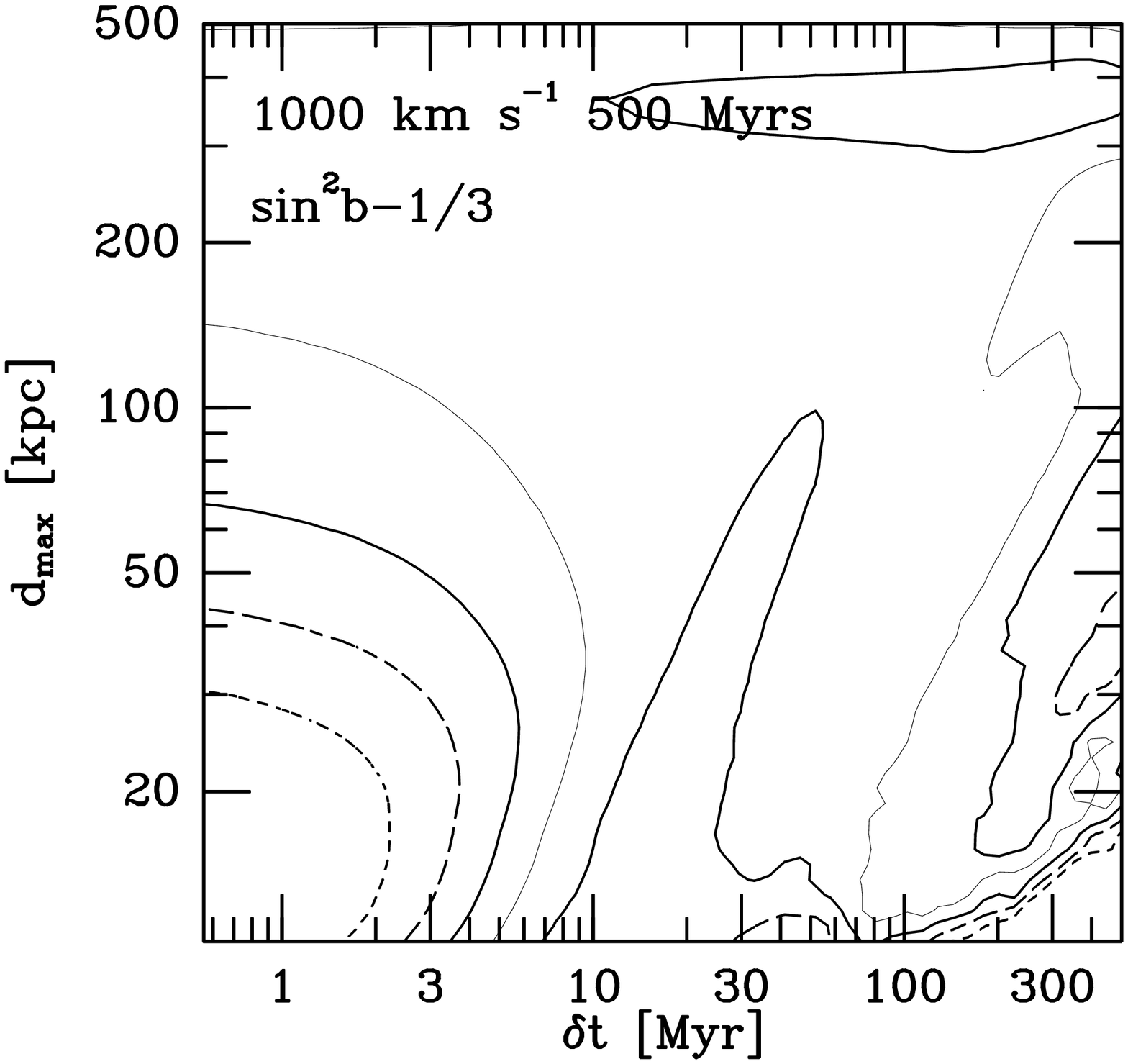,width=5.5cm,angle=-90}} \\
{\psfig{file=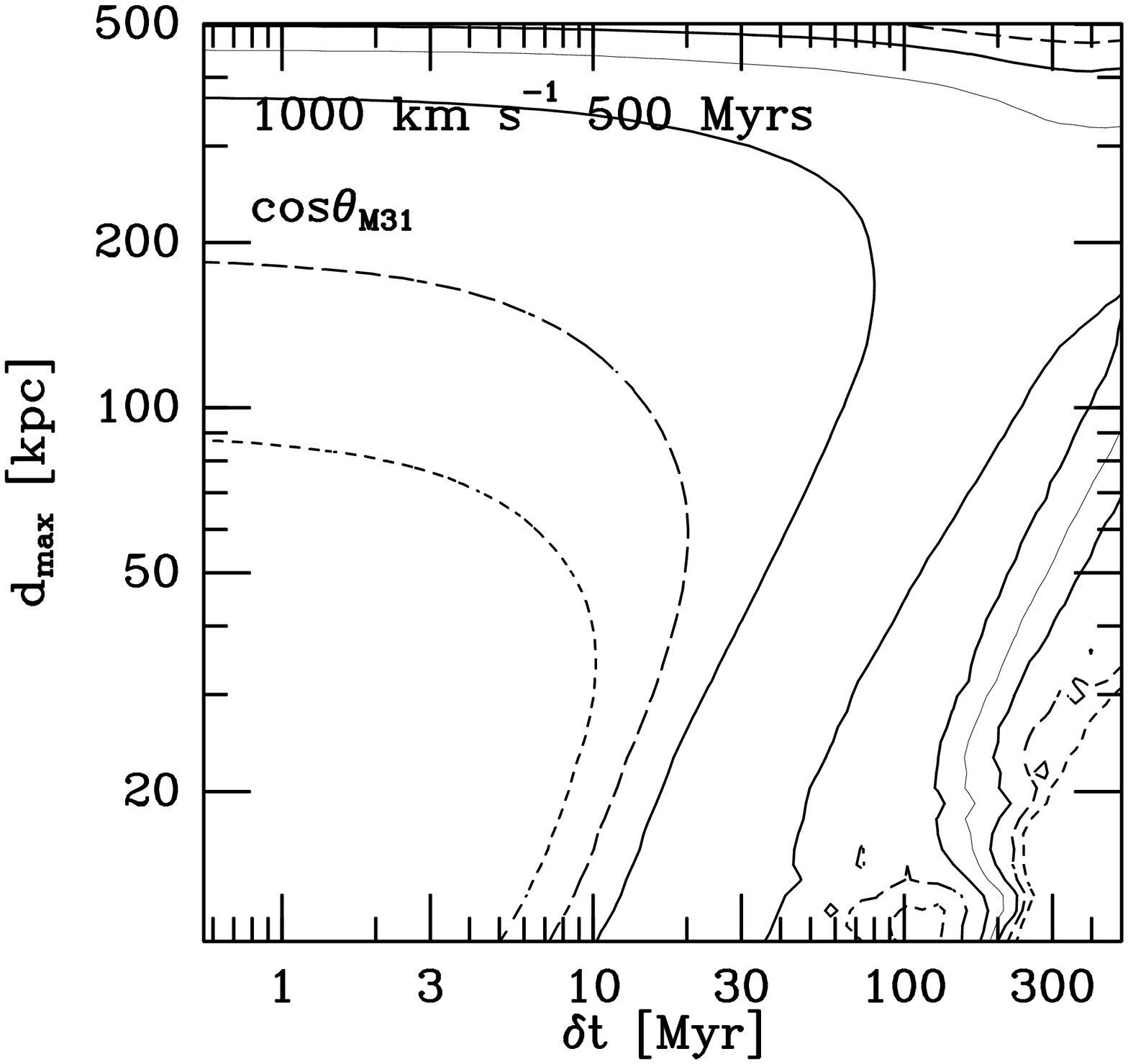,width=5.5cm,angle=-90}}  &
{\psfig{file=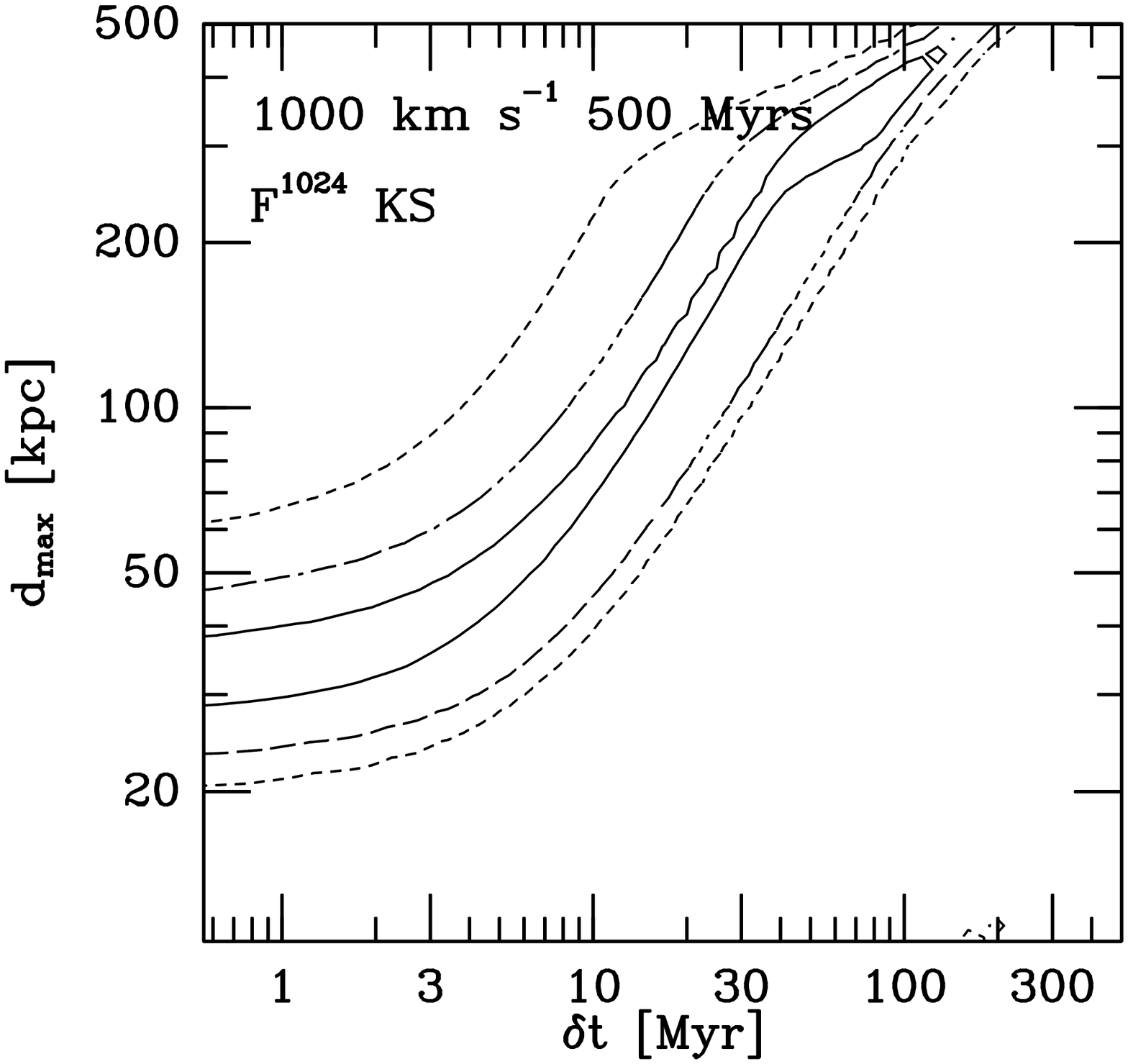,width=5.5cm,angle=-90}} \\
\end{tabular}
\end{center}
\caption{Comparison of a Galactic corona model with
the inclusion of M31 in which neutron stars
are born with a kick velocity of $1000$~\kms\ and have a burst-active
phase lasting $\Delta t = 500$ million years with a carefully-selected
sample of 285 bursts from the BATSE 2B catalogue.  Panels (a), (b), and (c)
show the contours in the ($\delta t$, $d_{\rm max}$)-plane along which
the Galactic dipole, Galactic
quadrupole moments, and dipole towards M31  of the model differ from
those of the data by $\pm$ 1$\sigma$ (solid lines), $\pm$ 2$\sigma$
(dashed line), and $\pm$ 3$\sigma$ (short-dashed line) where $\sigma$
is the model variance; the thin line in panel (a)-(c) shows the contour
where the dipole moment for the model equals that for the data.  Panel
(d) shows the contours in the ($\delta t$, $d_{\rm max}$)-plane along
which 32\%, 5\%, and $4 \times 10^{-3}$ of simulations of the
cumulative distribution of 285 bursts drawn from the peak flux
distribution of the model have KS deviations $D$ larger than that of
the data.
}
\vspace{-6mm}
\end{figure}

\subsection*{Sky and Brightness Distributions of Bursts}
\vspace*{-3.1mm}
Our detailed dynamical calculations of the  Galaxy show that a distant
corona of high velocity neutron stars can easily account for the
isotropic angular distribution and the brightness distribution of
GRBs (Figure 2) [see also \cite{LiDer:92,LTD:94,Pods:94}].

 \figside{\leavevmode\psfig{file=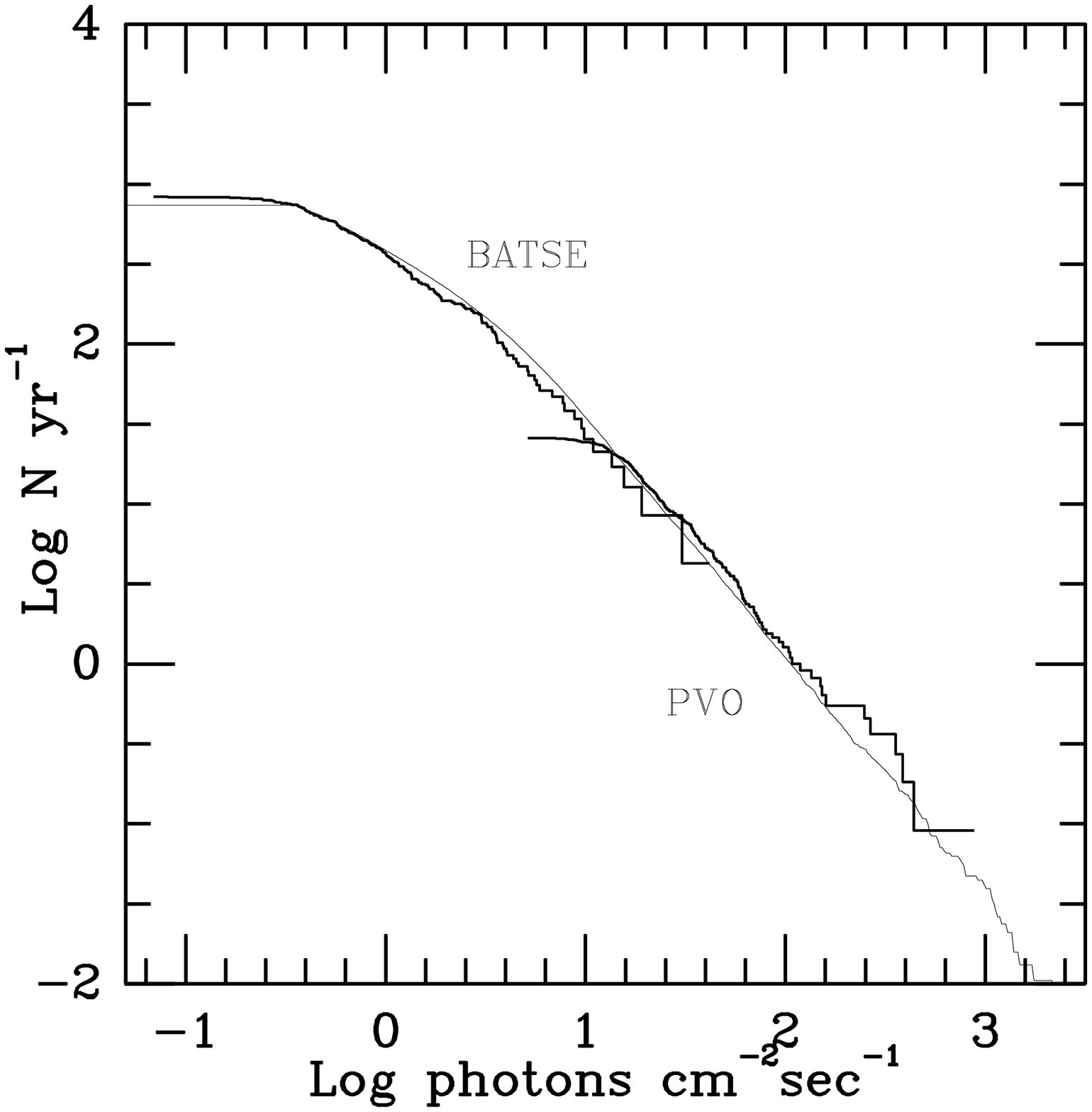,width=5cm,angle=-90}}
{\footnotesize{\bf FIG.~\ref{batse_pvo}}. \footnotesize
\baselineskip 1mm
Comparison of the brightness distribution of bursts from a Galactic
corona of high velocity neutron stars (thin line) and the brightness
distribution of both BATSE and PVO gamma-ray bursts (thick lines)
\cite{Fen93}.}{\label{batse_pvo}}

In high-velocity neutron star models, the slope of the cumulative peak
flux distribution for the brightest BATSE bursts and the PVO bursts
reflects the space density of the relatively small fraction of burst
sources in the vicinity of the Sun ($d \simless 50$ kpc).  A spread in
neutron star kick velocities, in neutron star ages at which bursting
behavior begins, or in the burst luminosity function tends to produce a
cumulative peak flux distribution with a slope of -3/2, the value
expected for a uniform spatial distribution of sources which emit
bursts that are ``standard candles.''  Figure~\ref{batse_pvo}
shows that a spread of
less than a factor of 10 in the luminosity function, which is
consistent with what know about GRBs,
is sufficient to produce agreement with not
only the BATSE, but also the PVO, brightness distribution of GRBs.
Beaming along the direction of motion of the
neutron star can also reproduce the combined BATSE and PVO brightness
distributions  \cite{DLT:93,LTD:94}.

The Galactic corona model predicts subtle anisotropies as a function of
burst brightness, which are a signature of the model and may offer a
means of verifying or rejecting it
\cite{LiDer:92,DLT:93,Pods:94,BulLam95a}.

\begin{figure}[th]
\vspace{-3mm}
\begin{center}
\begin{tabular}{lr}
{\psfig{file=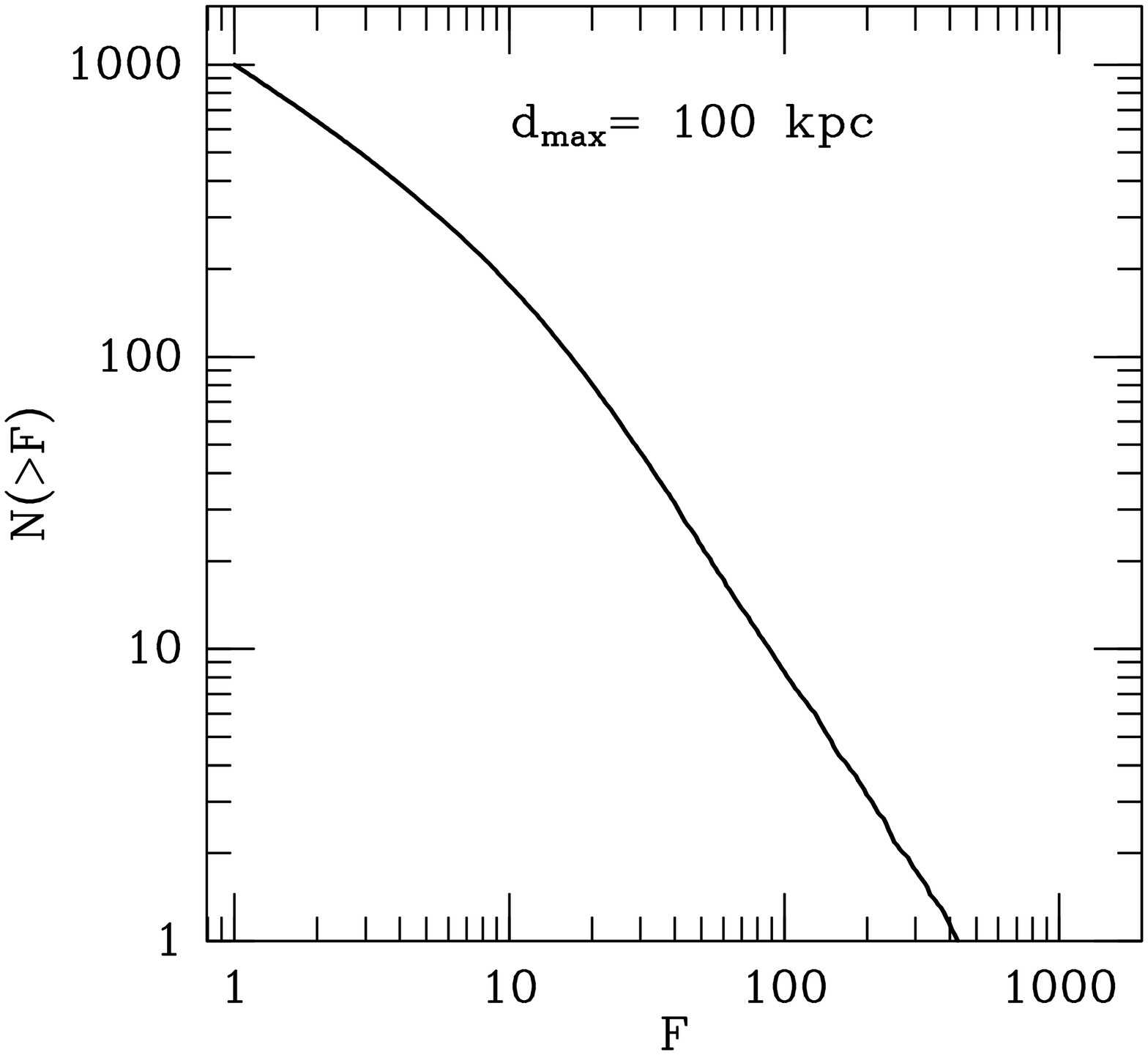,width=4.1cm,angle=-90}} &
{\psfig{file=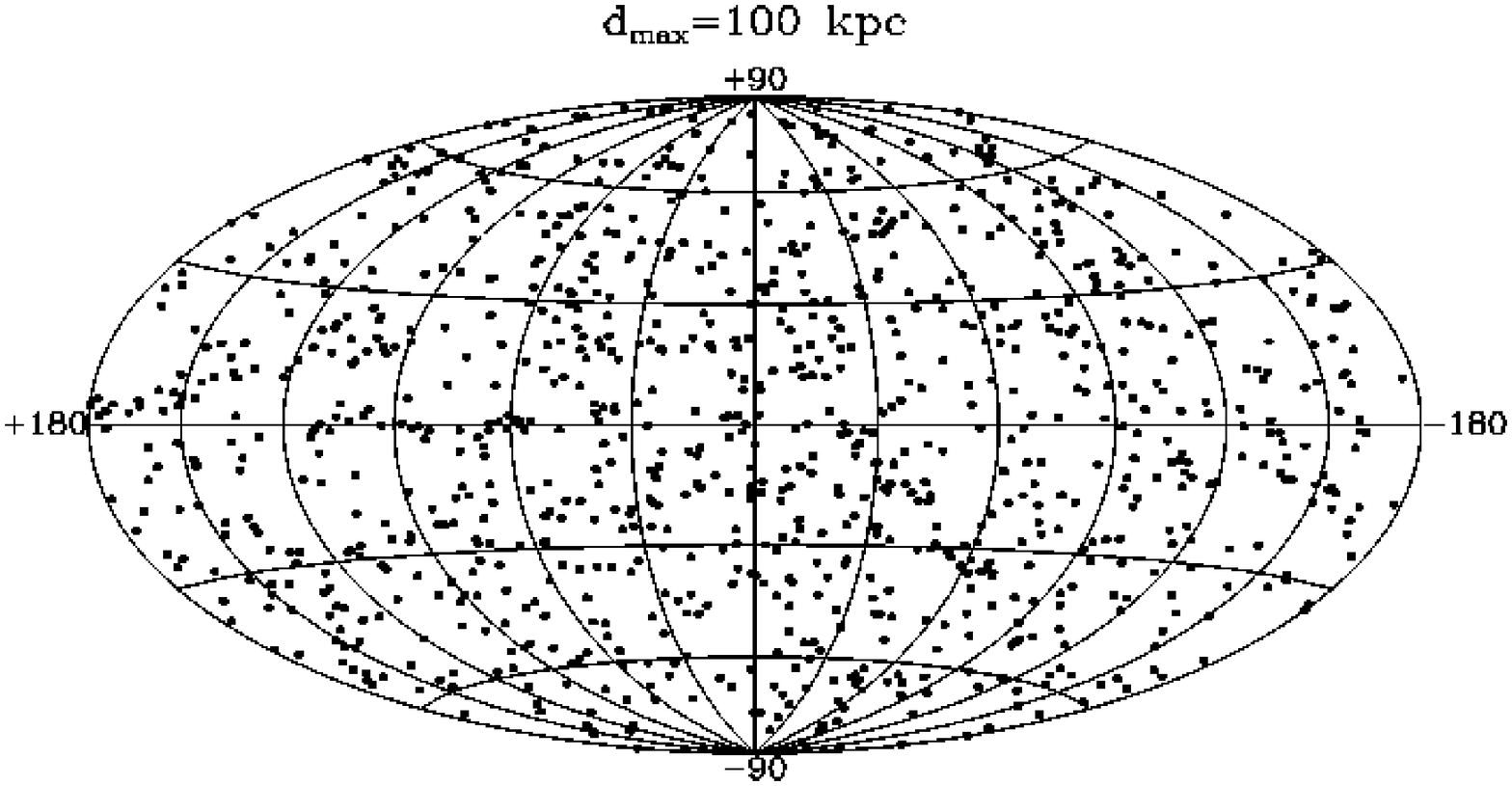,width=7.3cm,angle=-90}} \\
{\psfig{file=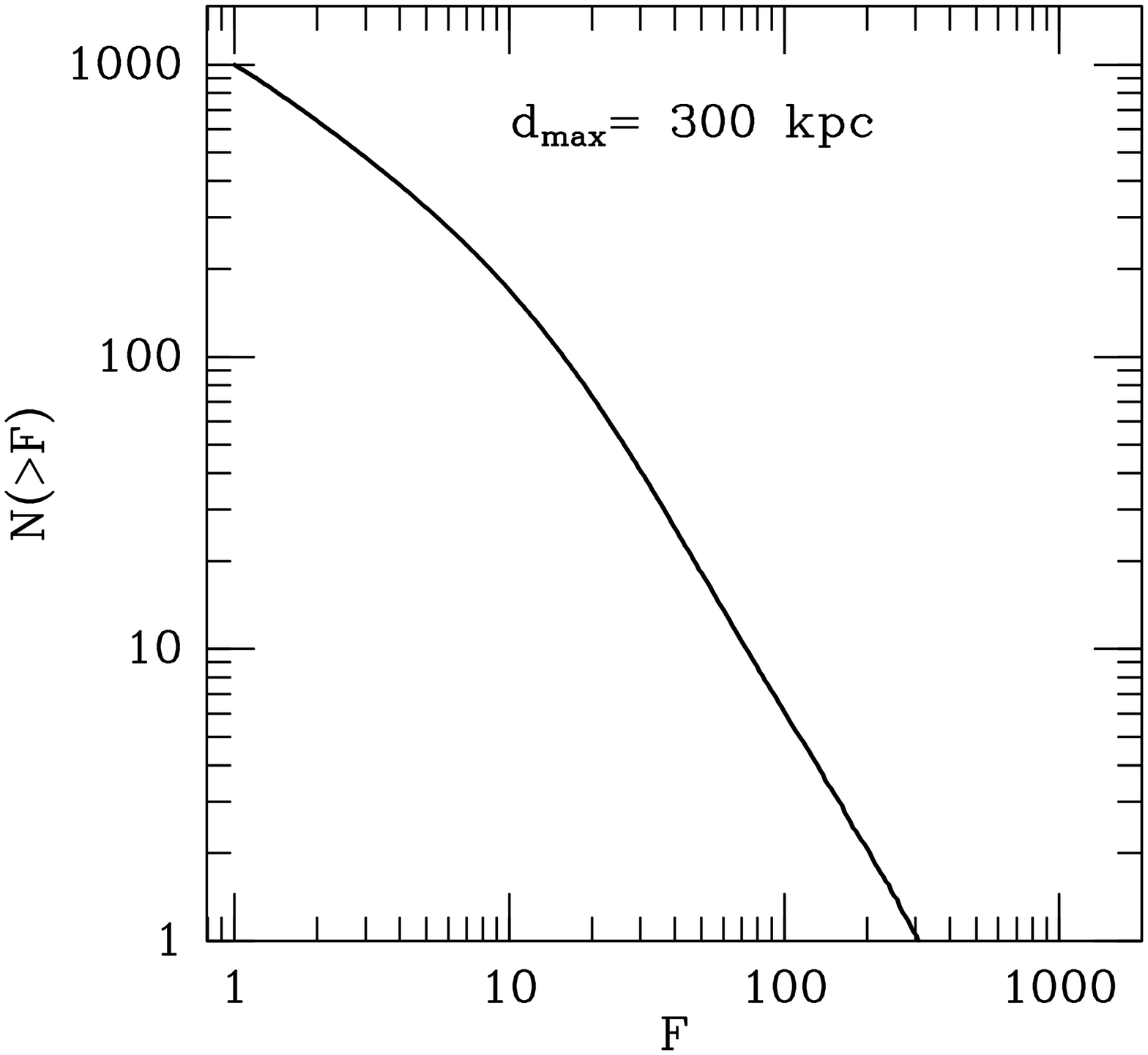,width=4.1cm,angle=-90}}  &
{\psfig{file=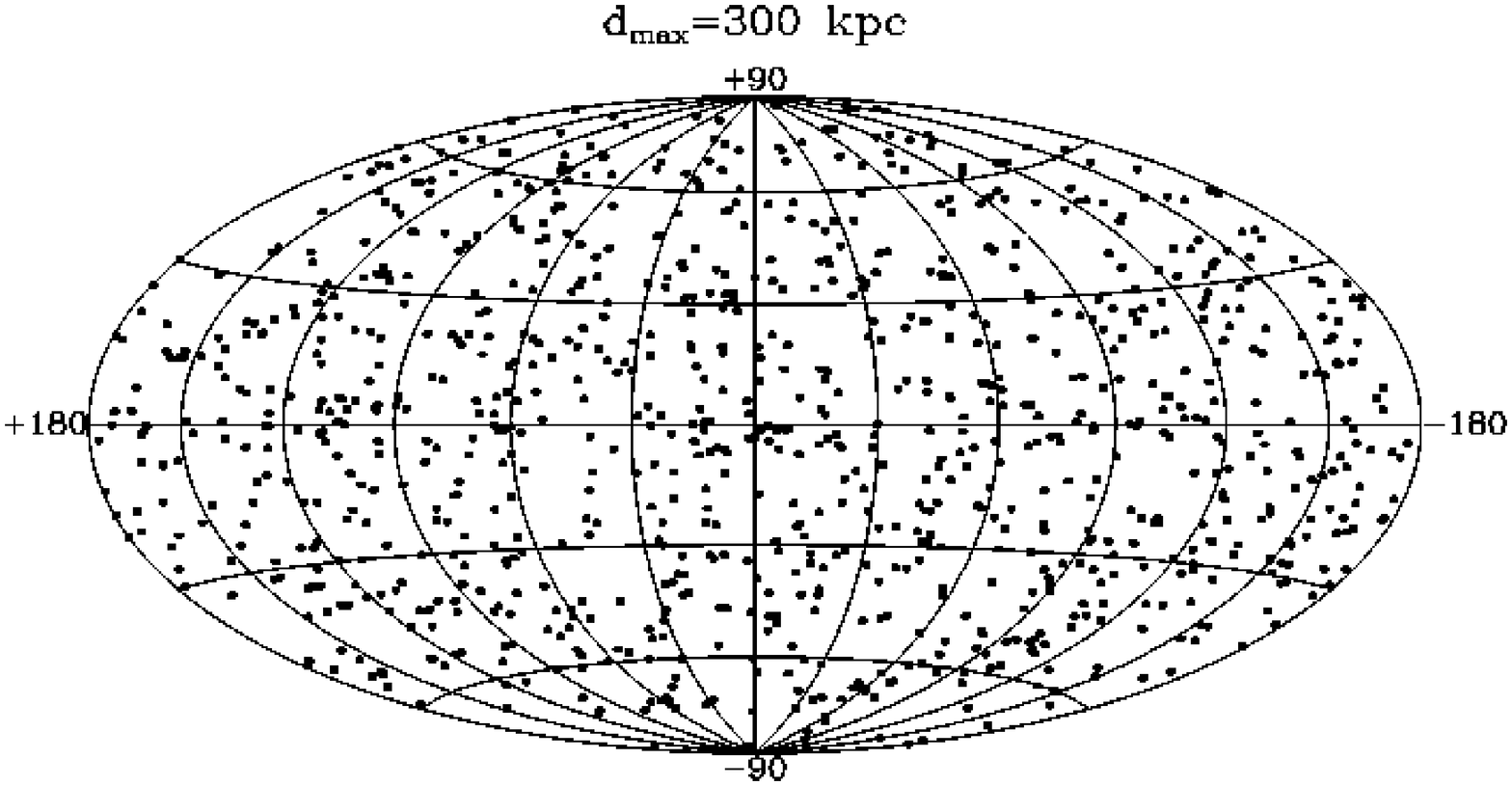,width=7.3cm,angle=-90}} \\
{\psfig{file=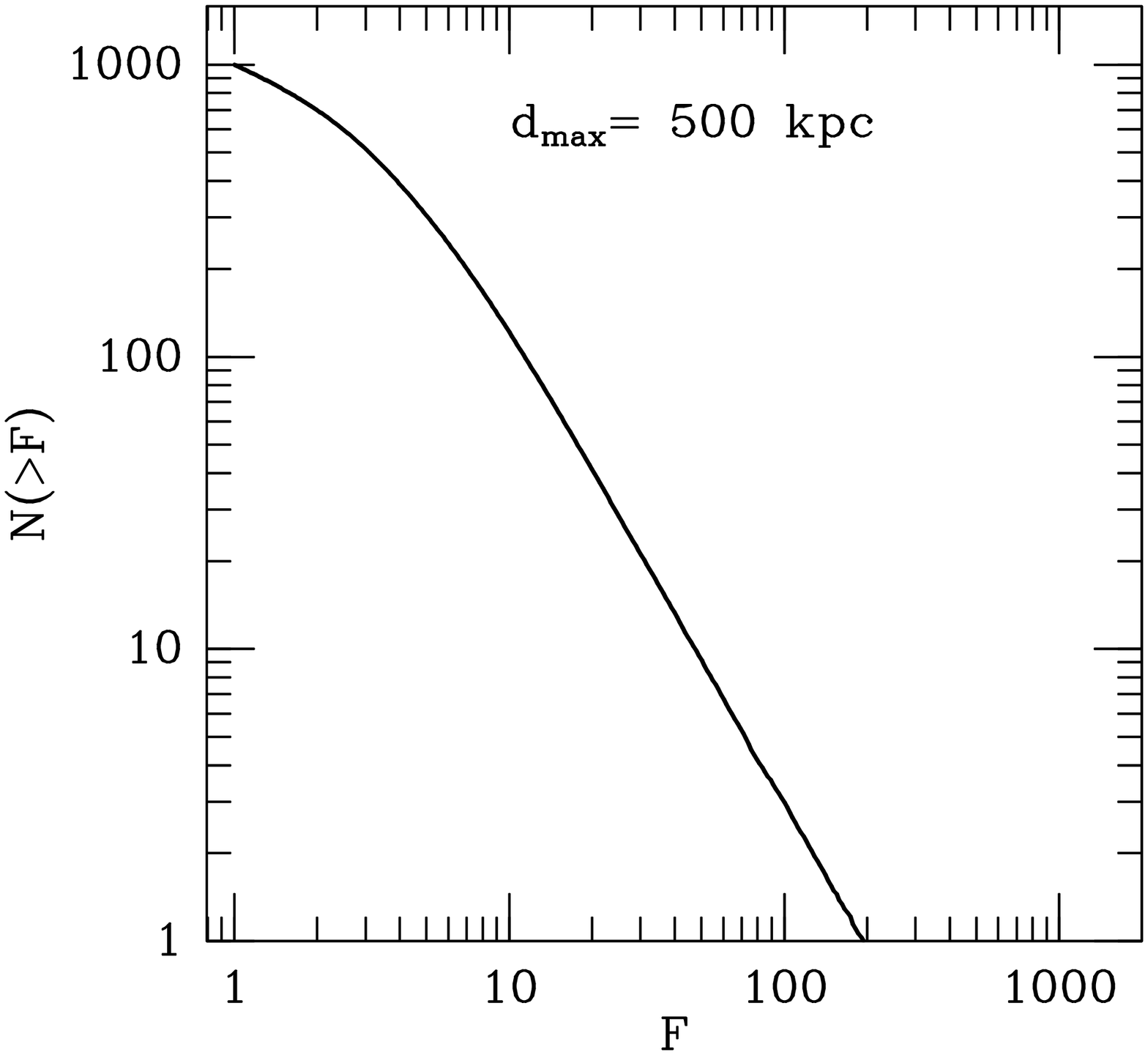,width=4.1cm,angle=-90}}  &
{\psfig{file=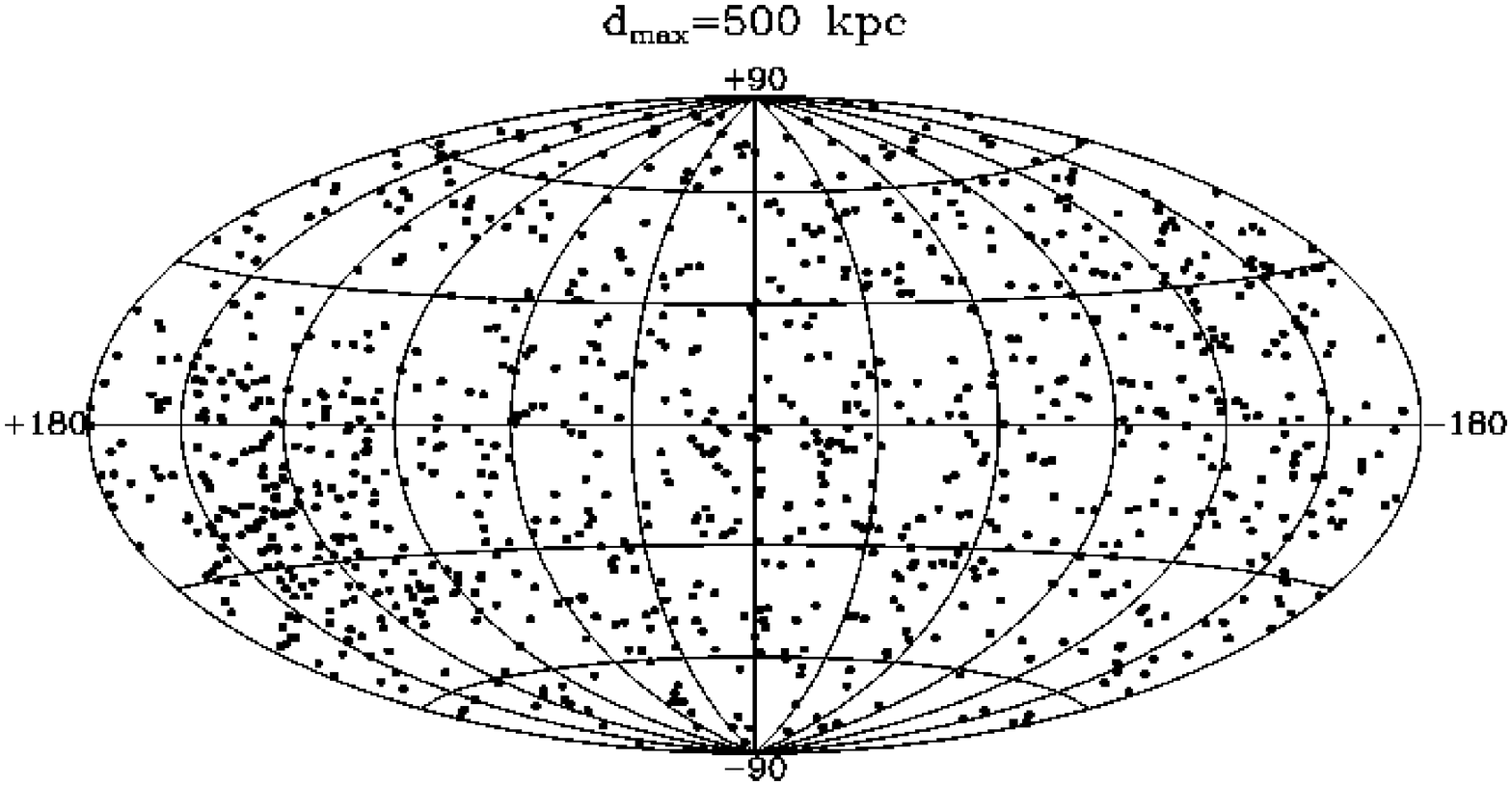,width=7.3cm,angle=-90}}
\end{tabular}
\end{center}
\vspace{-1.5mm}
\caption{Sky distribution and brightness distribution of bursts from
a Galactic corona of high velocity neutron stars for BATSE sampling
distances $d_{\rm max}$ = 100, 300, and 500 kpc. Note that an excess
of the bursts appears only when $d_{\rm max}$~=~500~kpc.}
\vspace{-7mm}
\end{figure}

It has often   stated that Andromeda, a bright galaxy similar to our
own Milky Way and lying only $700$ kpc away, imposes a severe
constraint on extended halo models \cite{Hak94}.
 This is true, however, only if the halo extends to large
distances \cite{BT87}.
 However, the halo of the
Milky Way can extend only 1/3 - 1/2 of the distance to Andromeda
because of tidal disruption.

A similar statement has been thought to be true for corona models
because in such models Andromeda produces its own ``wind'' of high
velocity neutron stars.  Some of these will travel toward us, and when
they produce GRBs, BATSE should detect them.

However, Andromeda imposes little constraint if the bursts are beamed
along the direction of motion of the neutron star, as some models posit
\cite{DLT:93,LTD:94}.
Then
only the rare neutron star in the corona of Andromeda whose motion is
almost directly toward or away from us would be visible.  So long as
the BATSE sampling depth $d_{\rm max} <$ 700 kpc (the distance to
Andromeda), the few bursts visible from Andromeda would always be
swamped by bursts from the many high velocity neutron stars born in the
Milky Way and moving away from us.  Only if $d_{\rm max} >$ 700 kpc, so
that a large number of the neutron stars in the Andromeda corona
whose motions are away from us are visible, would an excess toward
Andromeda be detectable \cite{BulCopLam:95c}. 

Even if the bursts radiate isotropically in all directions, detailed
dynamical calculations of the motion of neutron stars in the combined
gravitational potential of the Milky Way and Andromeda show that an
excess of bursts toward Andromeda is not detected until one samples
distances $d_{\rm max} \sim 500$ kpc from Earth (see Figure~4)
\cite{Pods:94,BulCopLam:95c}.
   Thus there is ample parameter space
(BATSE sampling distances $d_{\rm max} \approx 100 - 500$ kpc) for a
population of sources in a Galactic corona.

A larger sample of BATSE bursts or a more sensitive instrument might
reveal an excess of bursts toward Andromeda.
 If so, this would constitute definitive evidence that the
bursts are Galactic in origin.  Lack of an excess toward Andromeda
would be compelling evidence that the bursts are cosmological in origin
only if made by an instrument at least 50 times more sensitive than
BATSE, given the possibility that the bursts are beamed along the
direction of motion of the neutron star and current constraints on the
Galactic corona model.

\section*{Soft Gamma-Ray Repeaters}

\vspace*{-3.1mm}
We have seen that a Galactic corona of high velocity neutron stars can
easily account for the BATSE sky distribution and brightness
distribution of gamma-ray bursts.  Is there any evidence that high
velocity neutron stars can produce burst-like behavior?

Yes, there is.  Soft gamma-ray repeaters produce high energy transients
whose durations overlap with those of GRBs,
and whose characteristic spectral energies form a continuum with those
of GRBs.  The main distinction between SGRs and GRBs
is that the former have been
clearly shown to repeat on time scales of days to years
 whereas the latter have
been thought not to repeat.  But recently, a number of scientists have
found significant evidence that GRBs also repeat
\cite{QL93,WangLing93,WangLing95,Quashnock95}.

Three soft gamma-ray repeaters are known.  Two lie in the Galactic disk
at distances of tens of kpc (SGRs 1806-20 and 1900+14); the third lies
in in the Large Magellanic Cloud in the halo of the Milky Way at a
distance of 50 kpc.  All three are associated with young supernova
remnants
\cite{Evans80,Kulkarni93,Kouvel94,Murakami94,Hurley94}.
  In two cases, the
soft gamma-ray repeater lies far away from the center of the supernova
remnant, implying a neutron star velocity of $\simgreat 1000$ km
s$^{-1}$ \cite{Evans80,Hurley94}
Clearly, high
velocity neutron stars can produce burst-like behavior.

If GRBs come from high velocity neutron stars in a distant
Galactic corona, there are additional similarities between GRBs
and SGRs.  Both have luminosities $L \sim
10^{41}-10^{43}$ erg s$^{-1}$.  Both also appear to have strong
magnetic fields, as we discuss below.  These similarities and the ones
we discussed above suggest a physical or evolutionary relationship
between SGRs and GRBs.  The unification
of these two phenomena is a very attractive feature of the Galactic
hypothesis.

\section*{The Famous 1979 March 5 Gamma-Ray Transient}

\vspace*{-3.1mm}
We have seen that high velocity neutron stars can produce burst-like
behavior.  Have high velocity neutron stars ever been seen to produce
an event that looks like GRBs?  The answer is ``yes.''  The
event is the famous 1979 March 5 gamma-ray transient.

The source of this famous event is SGR 0526-66, which lies in in the
Large Magellanic Cloud in the halo of the Milky Way at a distance of 50
kpc.  It is associated with the young supernova remnant N49
\cite{Evans80,Rotsch94}
 SGR 0526-66 lies far away from the center of the supernova
remnant, implying a velocity greater than 1200 km s$^{-1}$.

Seventeen bursts have been observed from this source
\cite{Mazets79,Golen84}.  
 The distribution of the durations of
these bursts overlaps completely with that of GRBs.

 The burst had
an intense spike which lasted $\sim 0.2$ s, followed by $\sim 200$ s of
emission which exhibited an 8 s periodicity \cite{Mazets79}.
 The
association with the supernova remnant N49 and the 8 s periodicity
leave little doubt that this object is a neutron star.  The existence
of pulsations implies a strong magnetic field.  The spectrum of the
emission following the intense spike had a characteristic spectral
energy $\langle E \rangle \approx 40$ keV, typical of SGR bursts.

 Although nine different
satellites observed the March 5th event \cite{Evans80},
the intensity of the spike produced so-called ``dead-time'' and ``pulse
pike-up'' effects which precluded reliable analyses of the spectrum.
Recently, Fenimore et al. \cite{Feni95} used the power of present-day
computers to unravel these effects in the ICE and PVO instruments.
They found that the  spike has a characteristic spectral energy
$\langle E \rangle \approx 200$ keV, with no soft component,
like a typical gamma-ray burst.

Whether the 1979 March 5 event is a GRB or a unique event
can be argued either way.  But either way, it demonstrates that distant
high velocity neutron stars in the Galactic halo can produce events
that have the energy, the spectrum, and the duration of GRBs.
This evidence strongly supports the high velocity neutron star
model.

\section*{Energetics}

\vspace{-3.1mm}
   We take $F_{\rm peak} \sim 10^{-7}$ erg cm$^{-2}$ s$^{-1}$ as
the typical peak energy flux of a BATSE burst.  Then the typical burst
luminosity is
\begin{equation}
L_{\rm burst} \sim 10^{41}
\left({F_{\rm peak} \over {10^{-7} \; \erg \; \cm^{-2} \; \s^{-1}}}\right)
\left({d \over {100 \,\kpc}}\right)^2 \; \erg \; \s^{-1} \; .
\end{equation}
Taking 5 s as the average photon flux-to-energy flux conversion
factor for BATSE bursts \cite{Fen93}, the typical burst energy is
\begin{equation}
E_{\rm burst} \sim 5 \times 10^{41}
\left({F_{\rm peak} \over {10^{-7} \; \erg \; \cm^{-2} \; \s^{-1}}}\right)
\left({d \over {100\, \kpc}}\right)^2 \; \erg  \; .
\end{equation}

The rate of burst detection by BATSE corresponds to an all-sky rate
$R^{\rm BATSE}_{\rm burst} \sim 800$ bursts yr$^{-1}$.
Assuming a neutron star birth rate $R_{\rm NS} \sim 3 \times 10^{-2}$
yr$^{-1}$, each neutron star must produce a total number of bursts
\begin{equation}
N \approx {R^{\rm BATSE}_{\rm burst} /( {f_{\rm escape} R_{\rm NS}})}
\approx 8 \times 10^4 \left({{f_{\rm escape}} / {0.3}} \right)^{-1} \; ,
\end{equation}
during its burst-active phase, where $f_{\rm escape}$ is the fraction
of neutron stars born with velocities high enough to escape from the
Galaxy.  Then the total supply of energy needed by each neutron star in
the Galactic corona is
\begin{equation}
\label{etot}
E \sim N ({E_{\rm burst}/ {5 \times 10^{41}}}) \sim 10^{46} \; \erg \; ,
\end{equation}

Among possible energy sources are gravitational energy from accretion
of   planetesimals \cite{Cole95,Woosley95}, crustal strain energy
from spin down of the neutron star, and magnetic field energy stored in
the interior of the neutron star \cite{Pods:94}.  Below we discuss each
possibility in turn.

\subsection*{Accretion}
\vspace*{-3.1mm}
The gravitational energy released by accretion is
\begin{equation}
E_{\rm burst} = {{GM \,\Delta M_{burst}}/ R}\, .
\end{equation}
Taking a neutron star mass $M=1.4M_\sun$ and radius $10$~km,
\begin{equation}
\Delta M_{burst} \approx 10^{21} \left( {E_{\rm burst}/ {5 \times 10^{41} \;
\erg}} \right) \g \, .
\end{equation}
Then the total mass needed to power the GRBs from each neutron star is
\begin{equation}
M \approx N \Delta M \approx 10^{27} \left({E_{\rm burst} /
 {5 \times 10^{41} \; \erg}}\right) \g \approx 10^{-6} M_\sun\, .
\end{equation}
This amount of mass can be supplied by a planetesimal.

\subsection*{Crustal Strain Energy}
\vspace*{-3.1mm}
The rotational energy in a neutron star at birth is
\begin{equation}
E_\Omega \approx {1 \over 2} I \Omega^2 \approx 3 \times 10^{47} \left({P
/ {0.3\, \s}}\right)^{-2} \erg\, .
\end{equation}
However, only a fraction of this energy can be stored in the neutron star crust
and released at much later time is \cite{Pods:94}
\begin{equation}
E_{\rm strain} \approx 0.5 \, \mu \,\theta_{\rm max}^2 \,4\, \pi \, R^2\,
\Delta R_{\rm crust} \, ,
\end{equation}
where $\theta_{\rm max}$ is the maximum strain the crust can withstand before
braking, $R$ is the radius of the neutron star and $\Delta R$ is the thickness
of the crust.
Taking $\mu \approx 3 \times 10^{29}$ dyne cm$^{-2}$, $\theta_{\rm max}
\approx 10^{-2}$, and $\Delta R_{\rm crust} \approx 0.1 R \approx 10^5$
cm, the maximum strain energy that the crust can store is
\begin{equation}
\label{estrain}
E_{\rm strain} \approx 2 \times 10^{43} \hspace{-0.5mm}\left({\mu \over
{3\times 10^{29}\,
\dyne\, \cm^{-2}}}\right)\hspace{-0.5mm}\left({\theta_{\rm max} \over
10^{-2}}\right)^2 \hspace{-0.5mm}
\left({R \over {10^6\,
\cm}}\right)^2\hspace{-0.5mm}\left({{\Delta R_{\rm crust}} \over {0.1
R}}\right) \erg \; .
\end{equation}
This energy is much smaller that given by equation \ref{etot}.
  Thus the strain energy that can be
stored in the neutron star crust as it solidifies while the neutron
star is rotating rapidly appears unable to supply the total energy
needed to power the bursts in the Galactic corona model.

\subsection*{Magnetic Field Energy}
\vspace*{-3.1mm}
We know from accretion-powered pulsars and rotation-powered pulsars
that the surface fields of most neutron stars lie in the range $B_s
\sim 10^{11} - 10^{13}$ G.  We have virtually no knowledge about the
internal magnetic fields of neutron stars.  If the internal field
exceeds $10^{16}$ G, then superconductivity is quenched and the total
energy stored in the internal magnetic field is
\begin{equation}
E^{\rm normal}_{\rm magnetic} \approx {{4\pi} \over 3} R^3 {B^2 \over
{8\pi}} \approx {1 \over 6} R^3 B^2 \approx 10^{49} \left({B \over {10^{16}\,
\G}}\right)^2 \erg \; .
\end{equation}

If the internal magnetic field is less than $10^{16}$ G, it is expected
that the interior of the neutron star will be superconducting.  Then
the total energy stored in the internal magnetic field is \cite{Pods:94}
\begin{equation}
E^{\rm super}_{\rm magnetic} \approx {1 \over 6} R^3 B B_c \approx
10^{49} \left({B \over {6 \times 10^{13}\, \G}}\right)\left({B_c \over {10^{16}
\, G}}\right)
\erg \; .
\end{equation}

However, the energy stored in the interior magnetic field might then
 be released
on the time scale,
\begin{equation}
\tau_{\rm spindown} \approx 10^7 P^2 (B/10^{12}\, \G)^{-2}\, \yr \ll 5
\times 10^8 \, \yr \; ,
\end{equation}
if the spin vortices in the superfluid drag the magnetic flux tubes toward the
surface of the
neutron star\cite{Srinivasan90}.
Thus, if the interior of the neutron star is superconducting, the
amount of energy stored in the interior magnetic field is sufficient to
power the bursts in the Galactic corona model but may be released over
a period of time much less than the required lifetime of such
sources.

If magnetic field instabilities stress the neutron star crust, then
from equation \ref{estrain} above, the energy released would be
\cite{Pods:94}
\begin{equation}
E_{\rm strain} \approx 2 \times 10^{42} \left({\mu \over {3 \times 10^{29}
\,\dyne\, \cm^{-2}}}\right) \left({\theta \over 10^{-3}}\right)^2
\left({R \over {10^6
\, \cm}}\right)^2 \left({{\Delta R_{\rm crust}} \over {0.1 R}}\right) \erg \; ,
\end{equation}
which is about right for GRBs.

\section*{Radiative Processes}
\vspace{-3.1mm}
\subsection*{Pair Fireballs}

\begin{figure}[t]
\begin{center}
\begin{tabular}{lp{4cm}}
 \raisebox{-4.3cm}
{\leavevmode\psfig{file=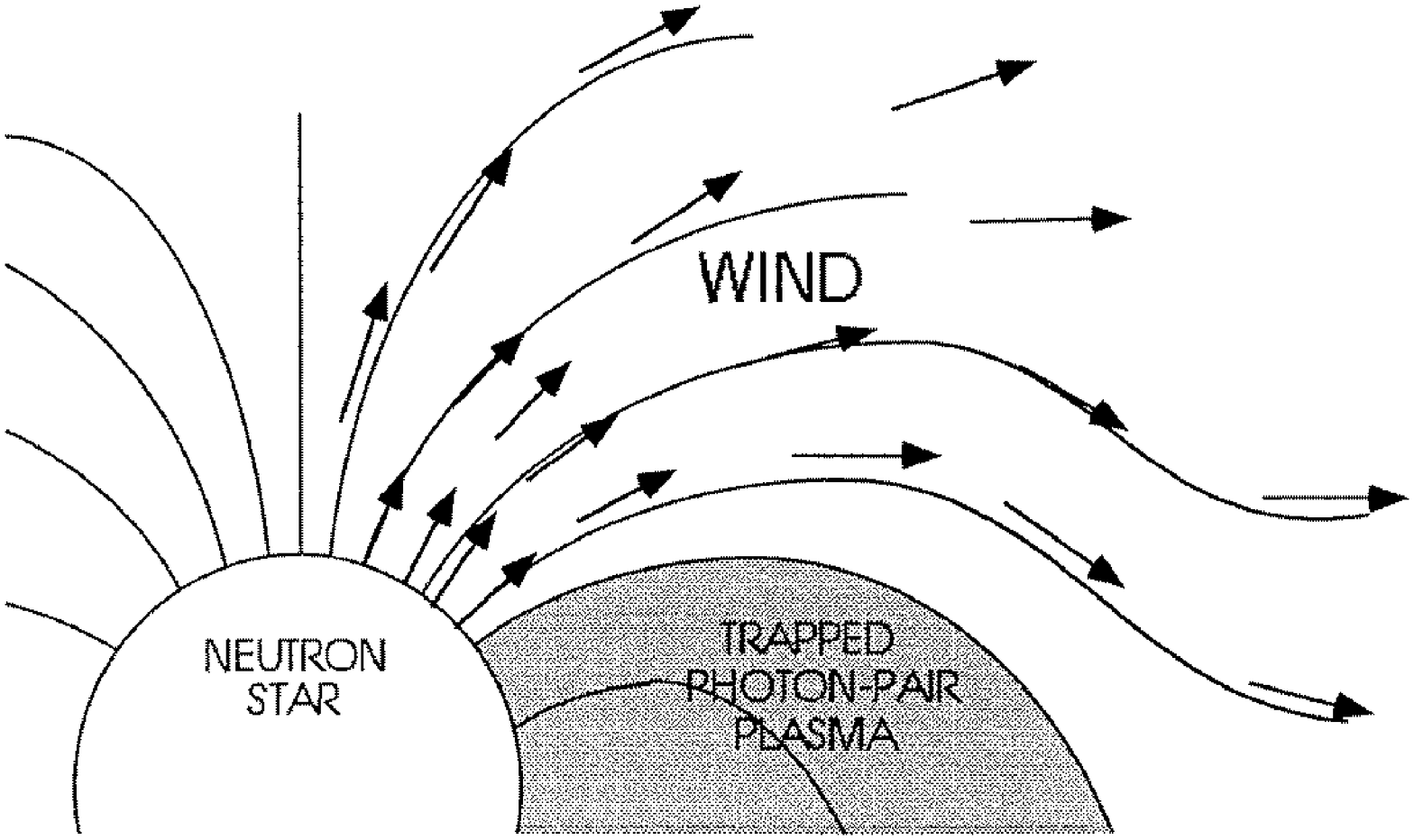,width=7cm,angle=-90}} &
{\footnotesize{\bf FIG.~\ref{windtb.ps}}
Neutron star magnetosphere, showing the region at the magnetic
pole where super-Eddington luminosities may drive a relativistic pair
wind, and the region at the magnetic equator where super-Eddington
luminosities may create a trapped pair fireball   \cite{TD95}.}
\end{tabular}
\end{center}
\refstepcounter{figure} \label{windtb.ps}
\vspace{-8mm}
\end{figure}

\vspace*{-3.1mm}
Our discussion of pair fireballs follows that of M\'esz\'aros \cite{Mesz95}.
A compactness parameter $
\tau_{\gamma \gamma} \approx ({{L_{burst} \sigma_T}  / {4 \pi r c
\epsilon_\gamma \Gamma^2}})\simless 1$
is required for photons to be observed above an energy
$\epsilon_\gamma$, where
$r$ is the radius of the source
and $\Gamma$ measures the relativistic velocity.  For $r \approx R$,
$\tau_{\gamma \gamma} \gg 1$ at
$\epsilon_\gamma \approx 1$ MeV unless $\Gamma \simgreat 10^4$.  This
result suggests that the initial stages of pair fireball in a gamma-ray
burst is optically thick.

If so, we expect that the sudden release of the energy that powers the
burst will produce a trapped fireball in the region of closed magnetic
fields lines in the neutron star magnetosphere and a mildly
relativistic wind from the regions of open field lines at the magnetic
poles (see Figure 5).  This picture is similar in many respects to the
soft gamma-ray repeater model of Thompson and Duncan \cite{TD95}.

The energy required in a solid angle $\theta$ is
\begin{equation}
E_{\rm burst} \approx 10^{41}\left({F_{\rm peak} \over {10^{-7} \,\erg\,
\cm^{-2}\, \s^{-1}}}\right) \left({d \over {100 \,\kpc}}\right) \theta^2 \,
\erg  \; .
\end{equation}

Pair production occurs if
$\epsilon_\gamma > 4 (m_e c^2)^2 \epsilon_t^{-1} \alpha^{-2} \; ,
$
where $\epsilon_t$ is the lab frame target photon energy and $\alpha$
is the relative angle between the two photons.

Causality implies $\alpha \simless \Gamma^{-1}$, or
$
\Gamma^{-1} \simless \alpha \simless 2 m_e c^2 (\epsilon_t
\epsilon_\gamma)^{-1/2} \; ,
$
and
\begin{equation}
\epsilon_\gamma \simless 10^4 (\epsilon_t/\MeV)^{-1} (\Gamma/10^2)^2
\MeV \; .
\end{equation}
This implies relativistic expansion and therefore beaming.  How large
might $\Gamma$ be?  In AGN, an initial value (near the central source)
as large as $\Gamma\sim 10^4$ is often assumed.  If the wind is powered by
Compton scattering, a value $\Gamma \approx 10$ or so is expected.

If $(E_{\rm burst}/\delta m_{\rm burst} c^2 )\simless 1$,
where $\delta M_{\rm burst}$ is the amount of baryonic matter entrained
in the outflow, then the wind will be subrelativistic due to baryonic
poisoning.

\subsection*{Cyclotron Lines}
\vspace*{-3.1mm}
Almost fifteen years ago
Mazets et al. \cite{Mazets81,Mazets82} reported seeing
single lines in the spectra of GRBs at low energies ($E
\simless 70$ keV).  Later Hueter \cite{Hueter88} reported single lines at low
energies in the spectra of two bursts seen by HEAO-1 A4.  However,
the statistical significance of the lines was modest.

More recently, equally-spaced lines were seen by {\it Ginga} in the
spectra of three bursts
\cite{Murakami88,Fenimore88,Graziani92,Yoshida92}
 with high significance \cite{Fenimore88,Graziani92,Freeman95}.
   The
line features in these three bursts have been studied extensively, and
there is no doubt that they exist.

Lines have not been definitively seen by BATSE \cite{Palmer94},
but this fact does not strongly contradict earlier observations
\cite{Band94}.

Similar line features are seen in the spectra of accretion-powered
pulsars\cite{MakiMih92},
which are known to
be magnetic neutron stars.  The equally-spaced lines seen in GRBs and in
accretion-powered pulsars are easily explained in terms
of cyclotron resonant scattering in a strong magnetic field
\cite{Wang89,B92}.

Magnetic neutron stars in the Galactic corona appear able to produce
cyclotron lines even though the luminosities of the bursts might greatly
exceed the so-called Eddington luminosity at which radiation pressure
and gravity balance.  Cyclotron lines may form, for example, in a
relativistic wind flowing out from the magnetic poles of the neutron
star \cite{Miller91}, 
or at the magnetic equator \cite{Freeman95}
 where hot plasma is trapped by the magnetic field
\cite{Lamb82,Katz82,Katz95,TD95}.

\section*{Conclusions}

\vspace{-3.1mm}
Detailed dynamical calculations   show that a distant Galactic
corona of high velocity neutron stars can easily account for the
isotropic angular distribution and the brightness distribution of
GRBs.
Gravitational potential energy from accretion and magnetic field energy
seem the most promising sources of energy which are
capable of powering bursts from such a population of neutron stars.

In a GRB hot plasma will likely flow from the polar cap in a relativistic wind,
but will be trapped at the magnetic equator by the magnetic field.
Cyclotron lines might form in either region.

\subsection*{Future Prospects}
\vspace*{-3.1mm}
Below we mention several key observations that might confirm or refute
the hypothesis that the GRBs come from a distant Galactic corona of
high velocity neutron stars.

{\bf Sky distribution.}  Our ability to detect or place upper
limits on any anisotropies in the burst sky distribution, especially as
a function of burst brightness, will increase slowly but steadily as
BATSE detects more bursts.  Confirmation of significant Galactic dipole
and/or quadrupole moments as a function of burst brightness, or
overall, would provide definitive evidence that the bursts are
Galactic.  Further limits on any angular anisotropy will constrain, and
might rule out, the Galactic hypothesis.  However, the limits that
BATSE will be able to achieve are not likely to be definitive, since
the angular distribution of bursts from the distant Galactic corona can
be very isotropic.

Detection of a concentration of bursts toward Andromeda, either by
BATSE, or by a
more sensitive   experiment
would constitute
definitive evidence that the bursts are Galactic in origin.  Lack of an
excess toward Andromeda would be compelling evidence that the bursts
are cosmological in origin only if made by an instrument at least 50
times more sensitive than BATSE, given the possibility that the bursts
are beamed along the direction of motion of the neutron star and
current constraints on the Galactic corona model

{\bf Cyclotron lines.}  Other spectroscopy instruments are now
operating (TGRS and Konus on Wind) or will soon be flown (e.g., HETE,
Konus on Spectrum X-Gamma, etc.) which will search for lines.  Further
confirmation of the existence of cyclotron lines would provide strong
evidence in favor of the Galactic hypothesis.

{\bf Repeating}.  The new bursts in the third BATSE catalog are not
expected to suffer from the same limitations which afflicted bursts in
the second year of observations due to failure of the tape recorders on
board the {\it Compton Gamma-Ray Observatory}.  It is therefore
expected that the third BATSE catalogue will provide an excellent
opportunity to test the repeating hypothesis.  Confirmation of
repeating would doom most cosmological models.

 \vspace{-3.1mm}


\end{document}